\documentclass[showpacs,floatfix,eqsecnum,aps,nofootinbib]{revtex4}
\usepackage{epsfig}

\begin{document}

\begin{flushright}
NSF-KITP-03-25
\end{flushright}

\title{Mirror dark matter and large scale structure.}
\author{A. Yu. Ignatiev}\email{sasha@physics.unimelb.edu.au}
\affiliation{School of Physics, Research Centre for High Energy
Physics, The University of Melbourne, Victoria 3010, Australia}
\author{R. R. Volkas}\email{r.volkas@physics.unimelb.edu.au}
\affiliation{School of Physics, Research Centre for High Energy
Physics, The University of Melbourne, Victoria 3010, Australia}
\affiliation{Kavli Institute of Theoretical Physics, University
of California, Santa Barbara, CA 93106-4030}

\begin{abstract}
Mirror matter is a dark matter candidate. In this paper, we re-examine
the linear regime of density perturbation growth in a universe
containing mirror dark matter. Taking adiabatic scale-invariant
perturbations as the input, we confirm that the resulting
processed power
spectrum is richer than for the more familiar cases of cold, warm
and hot dark matter. The new features include a maximum at
a certain scale $\lambda_{max}$, collisional damping below a smaller
characteristic scale $\lambda'_S$, with
oscillatory perturbations between the two.
These scales are functions of the
fundamental parameters of the theory. In particular, they decrease
for decreasing $x$, the ratio
of the mirror plasma temperature to that of the ordinary.
For $x \sim 0.2$, the scale $\lambda_{max}$ becomes
galactic. Mirror dark
matter therefore leads to bottom-up large scale structure formation,
similar to conventional cold dark matter, for
$x \stackrel{<}{\sim} 0.2$. Indeed, the smaller
the value of $x$, the closer mirror dark matter resembles
standard cold dark matter during the linear regime.
The differences pertain to scales smaller than $\lambda'_S$ in
the linear regime, and generally
in the non-linear regime because mirror dark matter is
chemically complex and to some extent dissipative.
Lyman-$\alpha$ forest data and the early reionisation epoch
established by WMAP may hold the key to distinguishing
mirror dark matter from WIMP-style cold dark matter.
\end{abstract}


\maketitle

\section{Introduction}

The dark matter problem provides one of the strongest reasons to
suspect the existence of physics beyond the standard model. We will
explore the possibility that dark matter is mirror matter in this paper.
Our objective is to understand the growth of density perturbations
in the linear regime in such a universe,
taking adiabatic scale-invariant perturbations as the input.
In doing so, we both confirm and extend the results of Ref.\ \cite{z}.
By comparing mirror dark matter (MDM)
with conventional cold, warm and hot dark matter (CDM, WDM and HDM, respectively),
we hope to explain the physics of mirror dark matter in as clear a way as possible,
and to pinpoint the data that are most sensitive to its characteristic
features.

Before launching into the analysis, we should set the stage by
briefly reviewing the evidence for non-baryonic dark matter,
and explaining why we think mirror matter is an interesting candidate.

It is very well established that the dynamics of
objects ranging in size from galaxies up to clusters of galaxies
cannot be understood using standard gravity unless one postulates
that invisible matter dominates over the visible by a factor of
$10-30$. (It is also logically possible that our understanding
of gravity at large scales is incomplete \cite{milgrom}, 
though we will not
pursue that possibility here.) It has been known for some time
that the conservative
option of the dark material being simply ordinary
matter in the form of non-luminous objects such as ``Jupiters'',
neutron stars, and so on, is ruled out if one accepts
that the light elements H, D, $^{3,4}$He and $^7$Li were created through
big bang nucleosynthesis (BBN). The baryon-to-photon ratio
is the one free parameter in standard BBN, and the value required
to produce fair agreement with the primordial abundance data is
a factor of five or so too small to account for all the dark matter
required to successfully model the gravitational dynamics of
clusters.\footnote{We prefer to add the electron-neutrino chemical
potential to the baryon-to-photon ratio in the parameter count of
BBN, with ``standard BBN'' then defined as the zero chemical potential
line in a two-dimensional parameter space. Some quite persistent
discrepancies in the data actually hint that a nonzero chemical
potential may be necessary \cite{bbn}.}
Ordinary baryonic dark matter is also
inconsistent with successful large scale structure formation,
principally because perturbation growth begins 
too late.\footnote{Large scale structure may be observationally
probed via galaxy surveys \cite{surveys} and
gravitational weak lensing \cite{weaklens}.}
In addition, acoustic peak data
from cosmic microwave background anisotropy measurements,
including those very recently reported by the
Wilkinson Microwave Anisotropy Probe (WMAP) collaboration,
have independently pointed to a matter-to-baryon density ratio of about
six \cite{wmap1, wmap2}.
So, if one accepts standard hot big bang cosmology, then
one must perforce accept the existence of non-baryonic dark matter.

Acknowledging the reality of non-baryonic dark matter, one can maintain
conservatism by supposing that massive ordinary neutrinos
provide the additional matter density. While this is a natural
and obvious possibility, it runs foul of large scale structure data.
A successful account of large scale structure, the concern of this paper and
one of the most important problems in physics \cite{p},
must be part of a
successful cosmology. Neutrino dark matter is the archetype of HDM,
inducing ``top-down'' structure formation whereby
very large structures form first, with smaller ones
arising from subsequent fragmentation. Hot dark matter driven
``top-down'' scenarios are now ruled out by the data: more
structure is observed at small scales than possible with
HDM (see, for instance, Refs.\ \cite{wmap2} and \cite{elgaroy}).

To sum up: the confluence of galactic/cluster dynamics,
big bang nucleosynthesis, acoustic peak,
gravitational lensing and large scale structure
data strongly point to a universe whose material or
positive-pressure component is
roughly $3\%$ luminous baryonic, $15\%$ dark baryonic, and $82\%$
exotic.
This is a remarkable conclusion.\footnote{Acoustic peak
and other data
also require the total matter density to be about $30\%$ of the
critical value giving a spatially flat universe \cite{wmap1, wmap2}.
With the strong
evidence for flatness from the acoustic peak features, one is also
obliged to add a $70\%$ non-material, negative-pressure
component called ``dark energy'', as also suggested by SN1a
observations \cite{sn1a}.
Note also that much of the dark baryonic matter, for high red shifts,
has now been detected through Lyman-$\alpha$ studies.}

It is very interesting that the exotic dark
component must consist of {\it stable} forms of matter, or at least
extremely long-lived. While exotic unstable particles abound in
extensions of the standard model, completely new stable degrees of freedom
pose a more profound model-building challenge. The stability challenge
is fully met by mirror matter.

The mirror matter model arose from the aesthetic desire to retain
the improper Lorentz transformations as exact invariances of nature
despite the $V-A$ character of weak
interactions \cite{ly, flv, blin}.\footnote{Alternative
motivations include E$_8\otimes$E$_8$ string theory, and brane-world
constructions such as the ``manyfold'' universe \cite{ah}.}
It does so
by postulating that, first, the gauge group of the world is a product
of two isomorphic factors, $G \otimes G$, and, second, that
an exact discrete $Z_2$ parity symmetry, unbroken by the vacuum,
interchanges the two sectors.\footnote{For the alternative of
spontaneously broken mirror symmetry see Refs.\ \cite{spon}.}
The minimal mirror
matter model takes $G$ to be simply the standard model gauge
group SU(3)$\otimes$SU(2)$\otimes$U(1). All ordinary
particles except the graviton receive a mirror partner.
A mirror particle has the same mass as the corresponding
ordinary particle, and mirror particles interact amongst
themselves in the same way that ordinary particles do, except
that mirror weak interactions are right-handed rather than
left-handed.

The two sectors must interact with each other gravitationally,
with certain other interaction channels also generally open, though
controlled by free parameters that can be arbitrarily small.
The possible non-gravitational interactions include photon--mirror-photon
kinetic mixing \cite{flv, pos}, neutrino--mirror-neutrino mass mixing \cite{flv2},
and Higgs-boson--mirror-Higgs-boson mixing \cite{flv, higgs}.
We will assume, for simplicity,
that all of these parameters are small enough to be neglected.
(One should bear in mind, however, that photon--mirror-photon kinetic
mixing can cause remarkable phenomena even with a controlling
parameter as small as $10^{-6}-10^{-9}$ \cite{pos, tung}.)

Mirror protons and mirror electrons are stable for exactly the
same reason that ordinary matter is stable. One would expect that
mirror matter would have existed in the cosmological plasma
of the early universe.
If so, mirror matter relics in the form of gas clouds, planets,
stars, galaxies and so on might well be common
in the universe today, manifesting observationally as dark
matter. Most prior work on MDM, with the notable exception
of Ref.\ \cite{z}, has focussed on the astrophysical phenomenology
of compact mirror matter objects, hybrid ordinary-mirror
systems, and diffuse mirror matter gas/dust in our
own solar system \cite{tung, mirrorastro}.
The discovery of mirror matter through such means
would, obviously, be a major breakthrough. In this paper,
however, we turn to the other generic purpose of
dark matter: to assist the growth of density perturbations
in the early universe, thus initiating large scale structure
formation. We want, ultimately, to know if mirror dark matter
is consistent with large scale structure data, and, if it is,
to develop observables that can discriminate between
MDM and the current paradigm of collisionless CDM (and whatever
other candidates might be dreamed up).

From the macroscopic perspective,
mirror matter is a much more complicated dark matter candidate
than standard CDM particles such as axions and WIMPs.\footnote{We
emphasise that the microscopic theory is by contrast very simple.} 
Rather
than just one species of particle, mirror dark matter
is chemically complicated,
consisting of all the mirror analogues of ordinary matter:
protons, neutrons and electrons. Further, MDM is
self-interacting, and a background of mirror
photons and mirror neutrinos interacts with the mirror-baryonic
matter. However, since (i) the self-interactions
of the mirror particles are by construction identical
to those of ordinary particles except for the chirality
flip, and (ii) the interaction between ordinary and mirror matter
is by assumption dominated by standard gravity,
the MDM universe can be analysed through well-defined physics
despite the complex nature of the dark sector.\footnote{Initial
conditions must also be supplied (see the next section).}
It is not
necessarily a virtue for DM to consist of a single exotic
species such as a WIMP or axion. Indeed, in their recent
review Peebles and Ratra \cite{peebles}
emphasised that standard CDM
can be viewed as the calculationally simplest DM scenario
that, in broad terms, is phenomenologically
acceptable, but which may be subject to revision
or replacement when more detailed large scale structure data
are collected. They then point out that certain data
already challenge standard CDM on points of detail, though
they caution that these discrepancies might
in the end be due to calculational problems only.
We take the view that all well-motivated standard model
extensions supplying stable exotic species should be
investigated for their DM potential.

The rest of this paper is structured as follows. In Sec.\ \ref{sect:MDMcosmo},
we review the elements of cosmology in a universe with
mirror matter, and explain why there need not be a $50/50$
mixture of ordinary and mirror matter. Sections \ref{sect:Jeans}
and \ref{sect:Silk} then discuss two key scales in the
perturbation evolution problem: the Jeans length
and Silk scale for mirror baryons. The former determines the
scale at which sub-horizon sized modes can begin to grow in
the dark matter (mirror) sector, while the latter determines
the scale below which growth is damped. Section \ref{sect:spectra}
discusses the outcomes of the linear growth regime through
final processed spectra for mirror dark matter perturbations.
It also discusses Lyman-$\alpha$ forest data, CMBR anisotropy
and early reionisation. We conclude in Sec.\ \ref{sect:conclusion}.

\section{Cosmology with mirror dark matter}
\label{sect:MDMcosmo}

We begin by dispelling a common misconception regarding mirror matter.
One might na\"{\i}vely expect
that the exact discrete symmetry between the ordinary and mirror
sectors in the Lagrangian
would require the universe to contain, and to have always contained,
a precisely $50/50$ mixture of ordinary and mirror particles. Such
a universe would be inconsistent with the standard cosmological
framework. First, the doubling of the universal expansion rate
due to mirror photons, neutrinos and antineutrinos would completely
spoil big bang nucleosynthesis. Even if some way could be
found to counteract the additional relativistic species\footnote{A coincident
epoch with a temporarily negative cosmological constant of the
right magnitude perhaps \cite{sorkin}?}, there would be a second objection.
As we reviewed above,
observations favour a DM to baryon density ratio of about five,
comfortably larger than two.

So, does an exact discrete mirror symmetry at the microscopic level
of fundamental interactions imply that the ordinary and mirror matter
densities in the universe must always be equal \cite{kst}?
In one sense, it is
trivial to see that the answer is ``No''. One may simply adopt
asymmetric initial conditions at the big bang. In that case,
the temperature $T'$ of the mirror plasma in the early universe,
and the background mirror photons today $T'_0$, is different
from that of the ordinary plasma, $T$, and the ordinary cosmic microwave
background photons today, $T_0$. One of the fundamental parameters
in our cosmology will therefore be
\begin{equation}
x \equiv \frac{T'_0}{T_0}.
\end{equation}
Since the energy density of relativistic species goes as the fourth
power of temperature, the contribution of the light mirror degrees
of freedom to the cosmological density is strongly suppressed by $x^4$.
Even a small difference between the temperatures, such as a factor of $1/2$,
is enough to comply with the BBN upper bound on extra relativistic energy
density. This removes the first objection to mirror matter cosmology.

One might be uncomfortable with ascribing the macroscopic asymmetry
of the universe to asymmetric initial conditions. However, standard
Friedmann-Robertson-Walker cosmology already has a raft of problems
associated with initial conditions: homogeneity, spatial flatness, etc.
One approach to those problems is, of course, inflationary cosmology.
Interestingly, prior work has shown that inflation can also effectively
initialise a mirror matter universe to have $T' \neq T$ \cite{kst, hodges}.
One way is
to introduce a mirror inflaton to partner the ordinary inflaton within
the chaotic inflation paradigm. Since the stochastic processes
germinating inflation will not
comply with the discrete symmetry on an event-by-event basis, the
main result follows.

What about the second objection to mirror matter cosmology? Given
that $T' < T$, one might conclude that the mirror baryons would
have a correspondingly lower density than their ordinary counterparts,
thus exacerbating the problem of not enough MDM. This conclusion
would be true if the magnitudes of the baryon asymmetries in the
two sectors were equal.

However, we have to take into account that the inequality of temperatures
of ordinary and mirror matter will in general change the outcome
of baryogenesis in the two sectors, even though the microphysics
is the same \cite{hodges}. One expects in fact that
\begin{equation}
\eta' \equiv \frac{n'_B}{n'_{\gamma}} \neq \frac{n_B}{n_{\gamma}}
\equiv \eta,
\end{equation}
where $n_X$ is the number density of ordinary species $X$ and
$n'_X$ is the number density of mirror species $X'$. (We denote
the mirror partner to a given particle by a prime.) The ratio
of mirror baryon and ordinary baryon number densities can be
written as
\begin{equation}
\frac{n'_B}{n_B} = \frac{\eta'}{\eta} x^3.
\end{equation}
Because the baryons and mirror baryons have equal rest masses and
are highly non-relativistic for
the epochs we consider, this quantity is also approximately the
energy density ratio,
\begin{equation}
\frac{\Omega'_B}{\Omega_B} \simeq \frac{\eta'}{\eta} x^3,
\label{eq:OmegaBratio}
\end{equation}
where, as usual, $\Omega_X$ denotes the energy density of $X$
in units of the critical density. In Refs.\ \cite{hodges, z},
it was shown that
the mirror baryon asymmetry can be {\it greater} than the
ordinary baryon asymmetry, and can in fact overwhelm the $x^3$
factor in Eq.\ (\ref{eq:OmegaBratio}):
\begin{equation}
\frac{\eta'}{\eta} > \frac{1}{x^3}.
\label{eq:etaprimeovereta}
\end{equation}
Two quite different
baryogenesis scenarios were analysed in Ref.\ \cite{z}: the
out-of-equilibrium baryon-number violating decays of massive bosons and
electroweak baryogenesis. Interestingly, in both cases
$\Omega'_B > \Omega_B$ with an acceptable $\Omega_B$ could be obtained
provided that
\begin{equation}
x \stackrel{>}{\sim} 0.01.
\label{eq:lowerlimitonx}
\end{equation}
We will use this value as an indicative lower limit to cosmologically
interesting values of $x$. It obviously should not be taken as
definitive, because we do not yet know what baryogenesis
mechanism actually operates in nature. (A new mechanism involving
mirror matter has been
very recently proposed in Ref. \cite{fv03}.)

Motivated by the above results, we take $\Omega'_B/\Omega_B$ as the
second free parameter in our cosmology, fixed only by observational
data.

Before moving on, we should deal with a third possible objection to
mirror dark matter, this one based on the results of recent works constraining
self-interacting CDM. Recall that an extension of standard CDM through
self-interactions was proposed to circumvent the problem of overdense cores for
some types of galaxies \cite{idm}.
The required properties were that the
elastic scattering cross-section of DM particle on DM particle should lie
in the interval $\sigma/M \simeq 10^{-23} - 10^{-24}$ cm$^2/$GeV, and
the DM should remain dissipationless. These constraints are violated by MDM
because it is dissipative and, if we take an atomic hydrogen cross-section
as a guide, then the self-interaction strength is too high. However, the
two cases are not directly comparable. The evolution of MDM is much more
complicated than that of the self-interacting CDM considered in
Ref.\ \cite{idm}.
For instance, MDM would form more intrinsic structure (mirror stars and
other compact objects) than self-interacting CDM, so it is not just a
question of scattering cross-sections. Exactly what sorts of compact
mirror matter objects would form, and how they would be distributed,
is a very complicated question beyond the scope of this work. This
development will {\it not} parallel that of the ordinary sector.
For instance, one of the key parameters affecting the galaxy formation
process -- the rate of star creation -- will be different because
primordial nucleosynthesis in the mirror sector will produce
much more mirror helium relative to mirror hydrogen than is the case
for their ordinary analogues \cite{z}.

Let us see how the main cosmological equations are changed by the
presence of mirror matter \cite{z}. The Friedmann equation for a flat
universe becomes
\begin{equation}
H^2 = \frac{8\pi G}{3}\rho_{tot},
\end{equation}
where $\rho_{tot}$ is the total energy density, ordinary
plus mirror (plus vacuum)
\begin{equation}
\rho_{tot} = \rho + \rho' + \rho_{\Lambda},
\end{equation}
and the Hubble parameter $H = \dot{a}/a$ where $a$ is the scale factor.
In terms of the present-day energy densities, $\Omega_{r,m,\Lambda}$
for radiation, matter and vacuum, respectively, and the present-day Hubble parameter
$H_0$ (the so-called Hubble constant), the Friedmann equation can be rewritten as
\begin{equation}
H(z)^2 = H_0^2\, [\, \Omega_r (1 + z)^4 +
\Omega_m (1 + z)^3 + \Omega_{\Lambda} \, ]
\end{equation}
where $z$ is red-shift.
(We include the vacuum energy contribution for completeness and
self-consistency only. Its effects are negligible for the
early universe epoch we will consider.)

The present energy density of
relativistic particles, $\rho_r$, is the sum of contributions from
ordinary photons, mirror photons, and presently relativistic neutrinos
and mirror neutrinos. It is given by
\begin{equation}
\rho_r = \frac{\pi^2}{30}\, [\, g_*(T_0) T_0^4 +
g'_*(T'_0) {T'_0}^4 \, ] = \frac{\pi^2}{30} g_*(T_0) T_0^4 (1 + x^4),
\label{eq:rhor}
\end{equation}
where $g_*(T_0)$ and $g'_*(T_0)$ are the effective numbers of relativistic
degrees of freedom in the ordinary and mirror sectors respectively:
\begin{equation}
g_*(T_0) = g'_*(T'_0) \simeq 2 (1 + 0.23 N^{rel\, \nu}_0).
\end{equation}
The number of presently non-relativistic neutrino flavours, $N^{rel\, \nu}_0$,
is either zero or one, depending on whether the neutrino masses are
degenerate or hierarchical, respectively.
From Eq.\ (\ref{eq:rhor}) we see that the contribution of mirror particles to
the relativistic energy density can be neglected at all times because
$x^4 \ll 1$ due to the BBN constraint $x \stackrel{<}{\sim} 0.5$.

Observations require us to take the mirror baryons to dominate the
total present-day matter density, viz.
\begin{equation}
\Omega_m = \Omega_B + \Omega'_B \simeq \Omega'_B.
\end{equation}
After WMAP, the favoured range at $68\%$ C.L.\ is
\begin{equation}
\Omega_m h^2 \simeq 0.14 \pm 0.02,
\end{equation}
where $h \simeq 0.72$ is the Hubble constant \cite{key}
in units of $\overline{H} \equiv 100$ km/s.Mpc.
For future convenience, we will use
\begin{equation}
y \equiv \frac{0.14}{\Omega_m h^2}
\end{equation}
instead of $\Omega'_B/\Omega_B$ as the second {\it a priori}
free parameter in our mirror matter cosmology. The $3\sigma$
preferred range for $y$ is $0.7 - 1.75$ \cite{wmap2}.

There are a number of the critical moments in the process of perturbation
growth. One of them
occurs when the universe is equally dominated by radiation and matter.
The corresponding redshift, $z_{eq}$, is found from
\begin{equation}
(\Omega_{\gamma} + \Omega'_{\gamma})(1 + z_{eq})^4
+ \frac{\rho_{rel\, \nu}(z_{eq})}{\rho_c}
= \Omega_m (1 + z_{eq})^3,
\label{eq:matterradequality}
\end{equation}
where $\rho_{rel\, \nu}(z)$ is the energy density in relativistic neutrinos
at matter-radiation equality.
Using Eq.\ (\ref{eq:rhor}), the observed present-day background photon
temperature and $x \ll 1$, this evaluates to
\begin{equation}
1 + z_{eq} \simeq \frac{5500}{y\xi},
\label{eq:zeq}
\end{equation}
where
\begin{equation}
\xi \equiv 1 + \frac{7}{8} \left( \frac{4}{11} \right)^{4/3} N^{rel\, \nu}
\simeq 1 + 0.23 N^{rel\, \nu},
\end{equation}
with $N^{rel\, \nu}$ denoting the number of relativistic neutrino flavours
at the designated moment. For the epoch prior to ordinary photon
decoupling, it is most likely that all three neutrino mass eigenstates are
always relativistic, so we will set $\xi \simeq 1.69$ in all of our
numerical estimates. Adopting this, Eq.\ (\ref{eq:zeq}) becomes
\begin{equation}
1 + z_{eq} \simeq \frac{3300}{y}
\end{equation}
yielding $z_{eq}$ in the range $1900-4600$ from the $3\sigma$
allowed interval for $y$.

Two other critical moments are matter-radiation decoupling in the
ordinary and mirror sectors. The exponential factor in the Saha equation
describing decoupling implies that these events occur at about
the same temperature \cite{z}, $T_{dec} \simeq T'_{dec}$, so that
\begin{equation}
1 + z'_{dec} \simeq \frac{1 + z_{dec}}{x} \simeq \frac{1100}{x}.
\label{eq:zdec}
\end{equation}
Matter-radiation decoupling in the mirror sector precedes that in
the ordinary sector because of the temperature hierarchy $x < 1$.
In the following, it turns out that we will have to consider two
cases defined by $x > x_{eq}$ and $x < x_{eq}$, where
\begin{equation}
x_{eq} \simeq 0.34 y.
\label{eq:xeqdefn}
\end{equation}
The distinction follows from Eqs.\ (\ref{eq:zeq}) and (\ref{eq:zdec}):
for $x > x_{eq}$, mirror radiation-matter decoupling occurs during the
matter-dominated epoch, while for $x < x_{eq}$ it occurs during the
radiation dominated epoch. Numerically, $x_{eq}$ takes values in the
approximate interval $0.24 - 0.6$ (the upper end of this range is
disfavoured by BBN).

\section{The Jeans length}
\label{sect:Jeans}

The Jeans length for mirror matter determines the minimum scale at
which sub-horizon sized perturbations in the
mirror matter will start to grow
through the gravitational instability in the matter-dominated epoch.
The mirror baryon perturbations begin growing first,
with the perturbations in the
ordinary matter catching up subsequently. This process is similar
to the standard CDM scenario, with the points of difference to be
discussed later.

Physically, the Jeans length sets the scale at which the gravitational
force starts to dominate the pressure force. It is defined as the
scale at which the sound travel time across a lump is equal to the
gravitational free-fall time inside the lump. The Jeans length
for mirror matter is given by
\begin{equation}
\lambda'_J(z) = \frac{\sqrt{\pi} v_s'(z) (1 + z)}{\sqrt{G \rho_{tot}(z)}},
\label{eq:Jeansdefn}
\end{equation}
where $v_s'(z)$ is the sound speed in the mirror matter, and the
$(1+z)$ factor translates the physical scale at the time of redshift
$(1+z)$ to the present time.

Let us examine the mirror matter Jeans length for the period
between mirror-neutrino decoupling and
mirror-photon decoupling. The sound speed is
calculated from
\begin{equation}
(v'_s)^2 = \frac{dp'}{d\rho'}
\end{equation}
where the pressure is dominated by the contribution from
mirror photons, $p' \simeq \rho'_{\gamma}/3$, and
the relevant density is given by
\begin{equation}
\rho' = \rho'_{\gamma} + \rho'_B.
\end{equation}
Using $\rho'_{\gamma} \sim {T'}^4$ and $\rho'_B \sim {T'}^3$,
we see that
\begin{equation}
(v'_s)^2 = \frac{1}{3}\frac{1}{1 + \frac{3}{4}\frac{\rho'_B}{\rho'_{\gamma}}}.
\label{eq:mirrorsoundspeed}
\end{equation}
To transform the sound speed expression into a more useful form,
we first note that
\begin{equation}
\frac{\rho'_B(z)}{\rho'_{\gamma}(z)} = \frac{\Omega'_B}{\Omega'_{\gamma}}
\frac{1}{1 + z}.
\end{equation}
Then, from the definition of $z_{eq}$ as given by Eq.\
(\ref{eq:matterradequality}),
we obtain that
\begin{equation}
\frac{\Omega_m}{\Omega_{\gamma}} \simeq (1 + z_{eq})(1 + x^4) \xi.
\end{equation}
Using $\Omega'_B = \Omega_m - \Omega_B$, we therefore deduce that
\begin{equation}
\frac{\rho'_B(z)}{\rho'_{\gamma}(z)}  = 
\xi \frac{1 + z_{eq}}{1 + z}
\left( \frac{1}{x^4} + 1 \right) \nonumber\\
 -  \frac{\Omega_B}{\Omega_{\gamma}}
\frac{1}{(1+z) x^4}.
\end{equation}
For our purposes it will be good enough to make the
approximations $\Omega'_B \simeq \Omega_m$ and $1/x^4 \gg 1$, so that
$\rho'_B(z)/\rho'_{\gamma}(z) \simeq
\xi (1 + z_{eq})/[x^4(1+z)]$. It is easy to check that
$3\rho'_B/4\rho'_{\gamma} \gg 1$ for the epoch of interest.
Using these approximations, Eq.\ (\ref{eq:mirrorsoundspeed}) yields
\begin{equation}
v'_s \simeq \frac{1}{\sqrt{3}} \frac{2 x^2 \xi}{\sqrt{3}}
\sqrt{\frac{1 + z}{1 + z_{eq}}}.
\label{eq:approxmirrorsoundspeed}
\end{equation}
Observe that the sound speed in the
mirror photon-baryon plasma is approximately proportional to $x^2$: the
low mirror sector temperature suppresses the sound speed by diluting
the relativistic component of the mirror plasma \cite{z}.

Substituting Eq.\ (\ref{eq:approxmirrorsoundspeed}) in Eq.\ (\ref{eq:Jeansdefn})
and evaluating the resulting expression we obtain
\begin{equation}
\lambda'_J(z) \simeq \frac{2.1 \times 10^4}{\sqrt{2 + z + z_{eq}}}\,
x^2 y^{1/2}\ {\rm Mpc},
\label{eq:Jeansatgeneralz}
\end{equation}
which implies that
\begin{equation}
\lambda'_J(z_{eq}) \sim 260 x^2 y\ {\rm Mpc},
\label{eq:Jeansatzeq}
\end{equation}
and
\begin{equation}
\lambda'_J(z'_{dec}) \sim
\sqrt{\frac{2 x}{x + x_{eq}}}\, \lambda'_J(z_{eq})\ {\rm Mpc}.
\label{eq:Jeansatzprimedec}
\end{equation}
Beware that if mirror photon decoupling occurs before matter-radiation
equality ($z'_{dec} > z_{eq}$, $x < x_{eq}$), then Eq.\ (\ref{eq:Jeansatzeq})
is inapplicable. The mirror baryon Jeans length plummets to very low
values after mirror photon decoupling, because the pressure supplied
by the relativistic component of the mirror plasma disappears, and
the sound speed greatly decreases.

\section{The Silk scale}
\label{sect:Silk}

For the case of mirror dark matter, perturbations on
scales smaller than a characteristic length $\lambda'_S$
will be washed out by the collisional or Silk damping
that arises while mirror photons decouple from the mirror
baryons \cite{silk}.
The microphysics of this process is identical to that of
Silk damping in a universe containing only ordinary baryons.
We use the latter (imaginary) universe as a familiar reference.

Elementary considerations involving photon diffusion may be used
to estimate that in our baryonic reference universe, the
Silk scale is given by
\begin{equation}
(\lambda^0_S)^2 = \frac{3}{5}
\frac{t^0_{dec} \lambda_{\gamma}(t^0_{dec})}{(a^0_{dec})^2},
\label{eq:Silkdefn}
\end{equation}
where $t^0_{dec}$ is the photon decoupling time, $\lambda_{\gamma}$
the photon mean free path, and $a^0_{dec}$ the scale factor
at decoupling. The mean free path is given by
\begin{equation}
\lambda_{\gamma} = \frac{1}{X_e n_e \sigma_T},
\end{equation}
where $X_e$ is the electron ionisation fraction at decoupling (so
that $X_e n_e$ is the total number density of free electrons) and
$\sigma_T = 8\pi\alpha^2/(3 m_e^2)$ is the Thomson scattering
cross-section. Using $n_e \sim n_B$, this expression
yields
\begin{equation}
\lambda_{\gamma} \sim \frac{G m_B m_e^2}{\overline{H}^2 \alpha^2}
\frac{1}{X_e}\frac{1}{(1 + z_{dec})^3}\frac{1}{\Omega_B h^2}.
\end{equation}
Using $X_e \sim 0.1$, $a^0_{dec} \simeq z_{dec} \sim 1100$,
and incorporating some refinements, one obtains the estimate
\begin{equation}
\lambda^0_S \sim 2.5 (\Omega_B h^2)^{-3/4}\ {\rm Mpc}
\end{equation}
for our reference baryonic universe \cite{kt}.

The mirror matter Silk length at mirror photon
decoupling can be obtained by deducing how
the various quantities in Eq.\ (\ref{eq:Silkdefn}) scale with
$x$. The results depend on whether mirror photon
decoupling occurs before or after matter-radiation
equality, i.e.\ whether $x < x_{eq}$ or $x > x_{eq}$,
respectively [see Eq.\ (\ref{eq:xeqdefn})].

\subsection{Case I: $x > x_{eq}$}

The sequence of events in this case is:
\begin{enumerate}
\item Matter-radiation equality occurs at redshift $z_{eq}$.
\item At a smaller redshift $z'_{dec}$ mirror photons decouple
from mirror baryons.
\item Later still, at $z = z_{dec}$, ordinary photon
decoupling occurs.
\end{enumerate}
After $z=z_{eq}$, perturbation growth is no longer
damped by the expansion rate, but it is still
retarded by mirror-photon
induced pressure. The latter disappears at $z=z'_{dec}$,
and perturbations above the scale $\lambda'_S$ (determined
below) begin to grow in the mirror sector. After $z = z_{dec}$,
ordinary photon pressure stops preventing the ordinary baryons
from falling into the potential wells created by the
now growing perturbations in the mirror baryons.

To evaluate the Silk scale, we note that
\begin{equation}
\frac{t'_{dec}}{t^0_{dec}} =
\frac{t'_{dec}}{t_{dec}} =
\left( \frac{a'_{dec}}{a_{dec}} \right)^{3/2} =
\left( \frac{1 + z_{dec}}{1 + z'_{dec}} \right)^{3/2} = x^{3/2},
\end{equation}
where $t_{dec}$ is the actual photon decoupling time.
The expression for $\lambda_{\gamma}$ tells us that
\begin{equation}
\frac{\lambda'_{\gamma}}{\lambda_{\gamma}} =
\left( \frac{1 + z_{dec}}{1 + z'_{dec}} \right)^3 = x^3.
\end{equation}
Hence,
\begin{equation}
\lambda'_S = x^{5/4} \lambda^0_S,
\end{equation}
where it is understood that $\Omega_B$ is replaced
by $\Omega_m$ in the expression for $\lambda^0_S$.
Observe that the mirror Silk scale is suppressed relative to
the reference universe analogue by the temperature
ratio to the stated power.

Numerically, we find that
\begin{equation}
\lambda'_S(x > x_{eq}) \sim 11 x^{5/4} y^{3/4}\ {\rm Mpc}.
\end{equation}

\subsection{Case II: $x < x_{eq}$}

In this case, the mirror photon decoupling has occurred
prior to matter-radiation equality, so perturbation
growth in the mirror sector begins at $z = z_{eq}$.
Scaling arguments that we will suppress establish that
the mirror Silk length is given by
\begin{equation}
\lambda'_S = \lambda^0_S x_{eq}^{-1/4} x^{3/2}
\end{equation}
for this situation. Once again, the temperature
ratio decreases the scale from that of the reference model.
Numerically,
\begin{equation}
\lambda'_S(x < x_{eq}) \sim 14 x^{3/2} y^{1/2}\ {\rm Mpc}.
\label{eq:mirrorSilkevaluated}
\end{equation}

\section{Processed power spectra for mirror dark matter}
\label{sect:spectra}

Having obtained the values of the characteristic scales for
our problem, we are now in a position to qualitatively
discuss the shapes of the MDM processed power spectra
in the linear regime, treating cases
I and II defined above separately.

\subsection{Processed power spectrum for case I: $x > x_{eq}$}

\begin {figure}[ht]
    \begin{center}
        \epsfxsize 12cm
        \begin{tabular}{rc}
            \vbox{\hbox{
$\displaystyle{ \, { } }$
               \hskip -0.1in \null} %\vskip 0.2in
} &
            \epsfbox{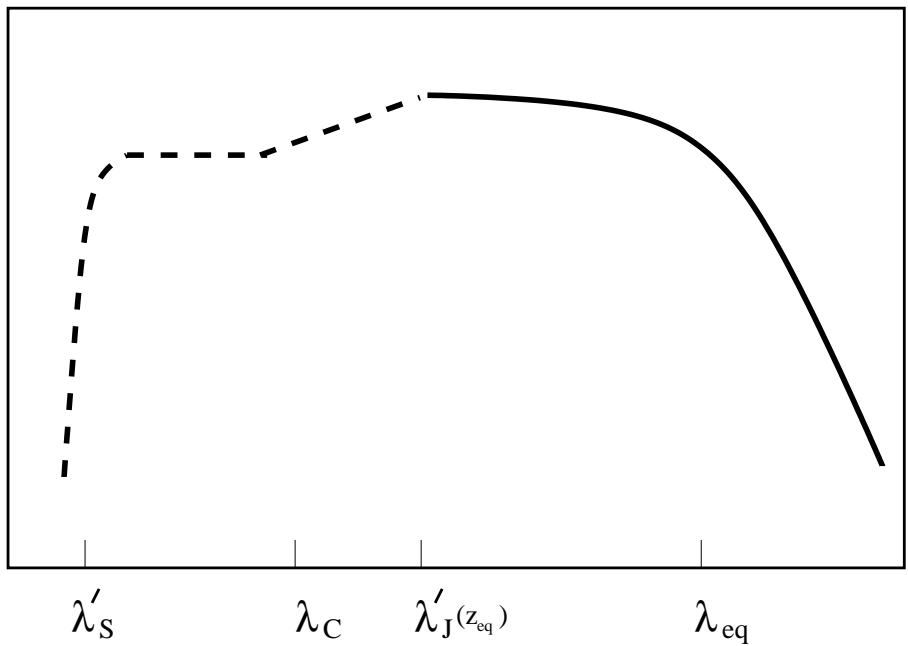} \\
            &
            \hbox{} \\
        \end{tabular}
    \end{center}
%\vskip 1in
\protect\caption
    {%
Schematic form of the processed power spectrum (not to scale) 
at $z=z_{eq}$ for case I ($x > x_{eq}$). The curve depicts
the mirror dark matter density perturbation $\log(\delta \rho'_B/\rho'_B)_{\lambda}$
as a function of the logarithm of the scale, $\log\lambda$. 
The dashed section represents
oscillatory evolution, while the solid line section shows
non-oscillatory evolution.
See the text for a full discussion
and for the defintion of the important scales indicated along the
horizontal axis.
 }
\label{fig1}
\end {figure}

\begin {figure}[ht]
    \begin{center}
        \epsfxsize 12cm
        \begin{tabular}{rc}
            \vbox{\hbox{
$\displaystyle{ \, { } }$
               \hskip -0.1in \null} %\vskip 0.2in
} &
            \epsfbox{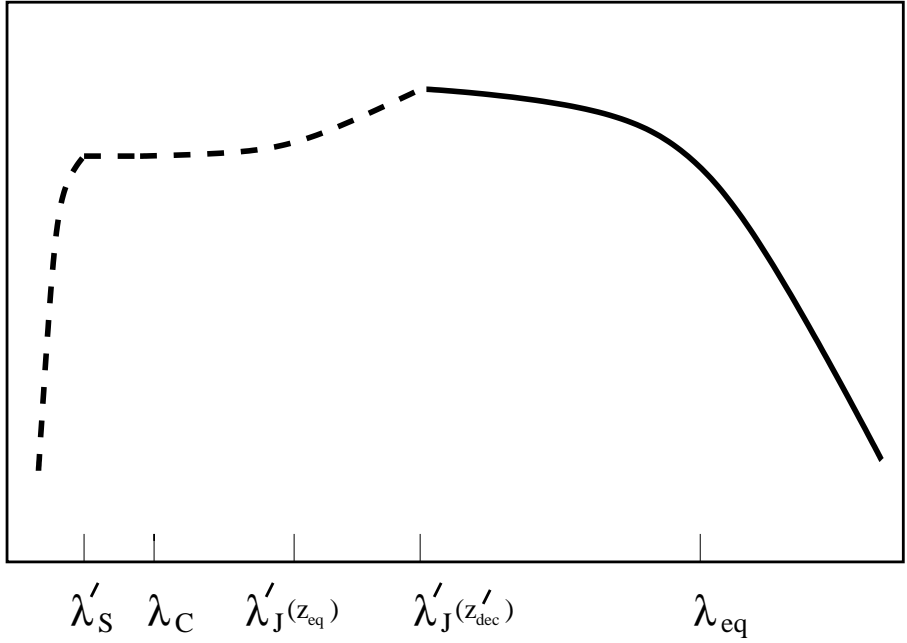} \\
            &
            \hbox{} \\
        \end{tabular}
    \end{center}
%\vskip 1in
\protect\caption
    {%
As for Fig.\ \ref{fig1}, except that $z = z'_{dec}$.
 }
\label{fig2}
\end {figure}

Figures \ref{fig1} and \ref{fig2} schematically depict the
processed power
spectrum in MDM when $x > x_{eq}$
at the two important moments of
matter radiation equality ($z = z_{eq}$) and
mirror photon decoupling ($z = z'_{dec}$), respectively.
The vertical axes are $\log(\delta \rho'_B/\rho'_B)_{\lambda}$,
the logarithm of the mirror baryon density perturbation
at scale $\lambda$,
at these two moments, plotted as functions of
$\log(\lambda/{\rm arbitrary\ scale})$. We will now explain how
these ``cartoons'' arise.

A number of scales are highlighted along the horizontal
axis. The largest of these is $\lambda_{eq}$, the
horizon scale at $z = z_{eq}$, which is easily
found to be given by
\begin{equation}
\lambda_{eq} \sim 130 y\ {\rm Mpc}.
\end{equation}
Scales smaller than $\lambda_{eq}$ have
already entered the horizon and have thus been processed by
gravitationally-driven growth and microphysical effects
such as Silk damping. Perturbations on super-horizon
sized scales obey
$(\delta\rho/\rho)_{\lambda} \sim \lambda^{-2}$ for
the usual reason, the prior period having been radiation
dominated.

The next scale is the mirror baryon Jeans mass, as
given by Eq.\ (\ref{eq:Jeansatzeq}). Notice that
the BBN bound $x \stackrel{<}{\sim} 0.5$ guarantees
that the mirror baryon Jeans scale is always
within the horizon at matter-radiation
equality. The
(sub-horizon sized) perturbations between
$\lambda_{eq}$ and $\lambda'_J(z_{eq})$ have grown
only logarithmically because of radiation
dominance.

The spectrum peaks at $\lambda'_J(z_{eq})$. Below that
scale, the perturbations are oscillatory and suppressed,
being washed out completely below the Silk scale
$\lambda'_S(z_{eq})$. To understand the detailed
behaviour in this interval, we need to look more
closely at the evolution of the perturbations.

Before doing so, however, we should pause to
note the similarities and differences between
Case I MDM and CDM. For scales $\lambda >
\lambda'_J(z_{eq})$, the processed spectrum
is identical to that of CDM.
Below $\lambda'_J(z_{eq})$, however,
the story is different. Recall that the CDM
spectrum maintains its slow logarithmic rise
as $\lambda$ decreases below $\lambda_{eq}$
until extremely small scales. This is because
both the Jeans and free-streaming lengths
for WIMP CDM are extremely small, the former
because there is no pressure support
to fight, and the latter because the WIMPs
are very massive and thus slow-moving.
For all practical purposes, CDM just has
two regimes ($\lambda$ larger or smaller
than $\lambda_{eq}$), while the physically richer
MDM has four.

We now turn to the relatively complicated
behaviour in the interval $\lambda'_S(z_{eq})
< \lambda < \lambda'_J(z_{eq})$. Consider
the time evolution of a perturbation at
scale $\lambda$ within this interval. A
relevant consideration is whether or not
$\lambda$ was smaller or larger than
the Jeans length at the $\lambda$
horizon-crossing time. One can compute that
\begin{equation}
z_{ent}(\lambda) \sim
\frac{4.6 \times 10^5}{\lambda/{\rm Mpc}},
\end{equation}
with Eq.\ (\ref{eq:Jeansatgeneralz}) then yielding
\begin{equation}
\lambda'_J(z_{ent}(\lambda)) \sim
\frac{360\, x^2 y}{\sqrt{1 + \frac{130 y}{\lambda/{\rm Mpc}}}}
\ {\rm Mpc}.
\end{equation}
The special scale $\lambda_C$ which is equal to the
Jeans length at $z_{ent}(\lambda_C)$ is then
\begin{equation}
\lambda_C \sim 140 y  \left( -1
+ \sqrt{1 + 6 x^4} \right)\ {\rm Mpc}.
\end{equation}
For $\lambda > \lambda_C$, the scale is larger than
$\lambda'_J(z_{ent}(\lambda))$, otherwise it is smaller.
Perturbations on scales $\lambda'_S < \lambda < \lambda_C$ therefore
start to oscillate about their horizon-entry values upon entering
inside the horizon; the averaged power spectrum is flat
within this interval.

Perturbations on scales $\lambda_C < \lambda < \lambda'_J(z_{eq})$
exhibit more complicated behaviour. Upon entering the horizon,
the perturbation begins to grow (logarithmically) slowly.
But the Jeans length increases as $z$ decreases, and at
some moment $z = z_{end}$ it overtakes $\lambda$. One
may estimate that
\begin{equation}
\frac{z_{ent}(\lambda)}{z_{end}(\lambda)} \sim
\frac{(\lambda/{\rm Mpc})\, y}{920 x^4 y^2
- 0.008 (\lambda/{\rm Mpc})^2}.
\end{equation}
The larger this ratio, the larger the perturbation growth.
It is easy to see that the ratio in fact increases monotonically
from $\lambda = \lambda_c$ until $\lambda = \lambda'_J(z_{eq})$,
which means that the growth factor for the perturbation also
grows in this interval. This completes the explanation of
the qualitative features of Fig.\ \ref{fig1}.

Figure \ref{fig2} depicts the processed spectrum at the
other critical moment: mirror photon decoupling at $z=z'_{dec}$.
After this moment, the mirror Jeans and Silk lengths fall to
very small values, and of course the universe is now matter
dominated. Perturbations at all scales therefore begin to grow
in proportion to the cosmological scale factor $a$,
the processed spectrum retaining its $z = z'_{dec}$ shape
(until linearity breaks down).

Many of the qualitative features of Fig.\ \ref{fig2} have the
same explanation as their counterparts in Fig.\ \ref{fig1}.
There are some differences, though.
Perturbations on scales larger than $\lambda'_J(z'_{dec})$
[see Eq.\ (\ref{eq:Jeansatzprimedec})] grow linearly
with $a$ because the universe is matter-dominated.
A scale between $\lambda'_J(z_{eq})$
and $\lambda'_J(z'_{dec})$ is larger than the Jeans
length at its horizon entry time, so the associated perturbation
grows until that scale is overtaken by $\lambda'_J$
before $z = z'_{dec}$.

The spectrum therefore has a maximum at the scale
\begin{equation}
\lambda_{max} = \lambda'_J(z'_{dec}).
\end{equation}
Using Eq.\ (\ref{eq:Jeansatzprimedec}), this evaluates
to
\begin{equation}
\lambda_{max} \sim 370
\frac{x^2 y}{\sqrt{1 + \frac{x_{eq}}{x}}}\ {\rm Mpc}.
\label{eq:lambdamax}
\end{equation}

\subsection{Processed power spectrum for case II: $x < x_{eq}$}

\begin {figure}[ht]
    \begin{center}
        \epsfxsize 12cm
        \begin{tabular}{rc}
            \vbox{\hbox{
$\displaystyle{ \, { } }$
               \hskip -0.1in \null} %\vskip 0.2in
} &
            \epsfbox{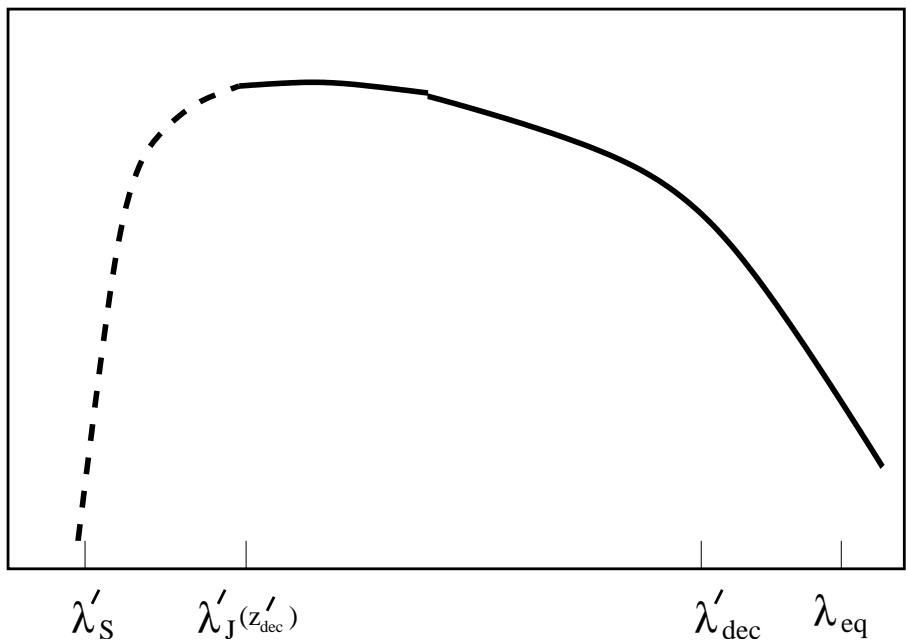} \\
            &
            \hbox{} \\
        \end{tabular}
    \end{center}
%\vskip 1in
\protect\caption
    {%
As for Fig.\ \ref{fig1}, except that it refers to case II ($x < x_{eq}$).
 }
\label{fig3}
\end {figure}

\begin {figure}[ht]
    \begin{center}
        \epsfxsize 12cm
        \begin{tabular}{rc}
            \vbox{\hbox{
$\displaystyle{ \, { } }$
               \hskip -0.1in \null} %\vskip 0.2in
} &
            \epsfbox{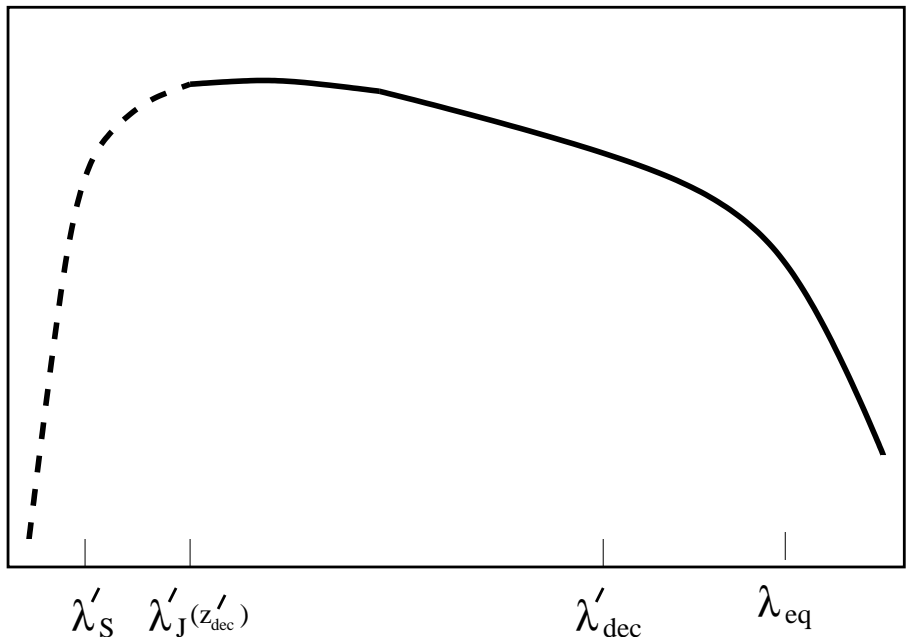} \\
            &
            \hbox{} \\
        \end{tabular}
    \end{center}
%\vskip 1in
\protect\caption
    {%
As for Fig.\ \ref{fig3}, except that $z = z_{eq}$.
 }
\label{fig4}
\end {figure}

For this case, mirror photon decoupling precedes matter-radiation
equality ($z'_{dec} > z_{eq}$). Figures \ref{fig3} and \ref{fig4} depict the
processed power spectra at these two moments. Their
qualitative features can be explained in a very similar
way to the preceding case. There are some points of
difference, however.

Examine Fig.\ \ref{fig3} first. It is similar to Fig.\ \ref{fig1},
but $\lambda'_{dec}$ (the horizon scale at $z = z'_{dec}$)\
plays the role previously held by $\lambda_{eq}$.
Between $\lambda'_{dec}$ and $\lambda'_J(z'_{dec})$, the
spectrum grows logarithmically slowly. Between
$\lambda'_J(z'_{dec})$ and $\lambda'_S(z'_{dec})$,
the spectral curve looks similar to the analogous
region in Fig.\ \ref{fig1}, except that there is no flat
part because the scale $\lambda_C$ is always
smaller than the Silk length: the curve is
growing, on average in this region, with
the perturbations also oscillating about the mean
for a given $\lambda$. The spectrum has a maximum
at the same scale as computed for case I [see
Eq.\ (\ref{eq:lambdamax})]. Once again, the
relatively quick fall off as the scale
decreases from $\lambda_{max}$ to the Silk
scale represents
qualitatively different behaviour compared
to standard CDM.

Turning to Fig.\ \ref{fig4}, the case II processed power spectrum at
matter-radiation equality, we see perturbations falling off as
$\lambda^{-2}$ for $\lambda > \lambda_{eq}$ and growing logarithmically
as $\lambda$ falls from $\lambda_{eq}$ to $\lambda'_J(z'_{dec})$. For
smaller scales, the form of the power spectrum remains unchanged from
its character at $z = z'_{dec}$, displaying slow logarithmic growth in
the direction of increasing $\lambda$. Recall that after mirror photon
decoupling, the mirror baryon Jeans length plummets to a very small
value, so all physically interesting scales are now greater than the
Jeans length and thus can grow. The growth imprinted on the processed
spectrum at $z = z_{eq}$ is, however, only logarithmic simply because
the preceding period was radiation dominated. The oscillations in this
regime created prior to $z = z'_{dec}$ remain as a feature of the
spectrum because there is no process that can damp them out. 
The position of
$\lambda_{max}$ thus remains unchanged.

For $z < z_{eq}$, the universe is matter-dominated
and perturbations on all scales grow
in proportion to the cosmological scale factor $a$.
The spectrum thus retains its shape at $z = z_{eq}$.

\subsection{Discussion}

\subsubsection{Linear regime}

We have seen that for both cases I and II, the
processed power spectrum at the relevant moment,
$z'_{dec}$ and $z_{eq}$ respectively, displays
a peak at $\lambda_{max}$ as given by
Eq.\ (\ref{eq:lambdamax}). For scales above
$\lambda_{max}$ the MDM perturbation spectrum
is identical to that of standard CDM.
At scales below the peak, the perturbations
oscillate about either a constant mean or one that
is slowly rising towards the peak.
Such oscillations are not a feature of
standard CDM.

The extent of the difference between MDM and
CDM therefore hinges on the value of
$\lambda_{max}$ (and $\lambda'_S$),
and hence on the
temperature ratio $x$. Furthermore,
the first structures to form will do
so at the scale $\lambda_{max}$.
It is therefore interesting to compare
$\lambda_{max}$ with the typical
galactic scale,\footnote{As
usual in this context, what we mean by this is
the size the material forming a galaxy would have
today had non-linearity not set in.}
\begin{equation}
\lambda_{gal} \sim 3.7 y^{1/3}\ {\rm Mpc}.
\end{equation}
Putting $y=1$ we see that $\lambda_{max}$
is equal to the galactic scale at
\begin{equation}
x \sim 0.2.
\end{equation}
For $x > 0.2$, the first structures to form
would be larger than galaxies, while for
$x < 0.2$ the scale falls rapidly into
the sub-galactic regime.

If one classifies dark matter as CDM-like,
WDM-like and HDM-like according to whether the
first structures are sub-galactic, galactic
or super-galactic, respectively, then we conclude
that MDM is
\begin{enumerate}
\item CDM-like for $x \stackrel{<}{\sim} 0.2$,
\item WDM-like for $x \sim 0.2$, and
\item HDM-like for $x \stackrel{>}{\sim} 0.2$.
\end{enumerate}
Given that top-down structure formation
appears to be disfavoured, we conclude that
\begin{equation}
x \stackrel{<}{\sim} 0.2
\end{equation}
is the favoured temperature range for
large scale structure formation.

The smaller $x$ is, the more closely the
MDM processed power spectrum in the linear regime
resembles its analogue for standard CDM.
It is interesting that MDM becomes CDM-like for
$x$'s that are not too small relative to our
indicative lower limit of about $0.01$ [recall the
discussion leading to Eq.\ (\ref{eq:lowerlimitonx})].

A stringent test of the perturbation spectrum
at small scales in the linear regime arises from
Lyman-$\alpha$ forest data \cite{weinberg}. 
A detailed study
of the implications of these data for MDM is
certainly well motivated. In this paper we will
have to settle for the rough guidance provided
by existing analyses constraining warm dark matter.
MDM and WDM are similar in that each exhibits
small scale wash-out, though the mechanisms are
different (Silk damping and free-streaming,
respectively). In Ref.\ \cite{narayanan}, Narayanan et al
use Lyman-$\alpha$ data to
constrain the mass $m_{WDM}$ of the WDM particle
through its free-streaming scale $R$,
given by
\begin{equation}
R \simeq 0.2 (\Omega_{WDM} h^2)^{1/3}
\left( \frac{m_{WMD}}{{\rm keV}} \right)^{-4/3}\ {\rm Mpc}
\label{eq:WDMfreestreamingscale}
\end{equation}
in terms of the cosmological WDM density $\Omega_{WDM}$.
They obtain
\begin{equation}
m_{WDM} > 0.75\ {\rm keV}
\label{eq:WDMmassbound}
\end{equation}
for an $\Omega_{WDM} h^2 = 0.2$ universe. Adopting the
resulting upper bound on $R$ as a rough bound on
$\lambda'_S$, we get from Eqs.\ (\ref{eq:mirrorSilkevaluated}),
(\ref{eq:WDMfreestreamingscale}) and (\ref{eq:WDMmassbound})
that
\begin{equation}
x \stackrel{<}{\sim} 0.06,
\end{equation}
where $y = 0.14/0.2$ was used for consistency. This bound appears to be
more stringent than the $x < 0.2$ range obtained by requiring that
$\lambda_{max}$ be sub-galactic.

\subsubsection{CMBR acoustic peaks}

Over the past decade the measurements of the Cosmic Microwave Background
anisotropy provided us with a powerful tool for finding the cosmological
parameters and studying the structure formation in the Universe. In
particular, recent WMAP data \cite{wmap1,wmap2} have established that
the adiabatic perturbation scenario is favoured over isocurvature
perturbations. A natural question is then: could the CMB data help us in
discriminating between MDM and other types of dark matter?

From our previous discussion it follows that the new features of the MDM
power spectrum might leave an imprint on the CMB anisotropy at small
angular scales, dictated by Eq.(\ref{eq:lambdamax}). With decreasing
$x$ these angular scales become smaller which corresponds to the larger
values of $\ell \sim \theta^{-1}$. Thus we could expect a non-negligible
effect only at larger $x$ which is comparatively less interesting in
view of the arguments presented above. However, it could be worthwhile
to give further precision to this qualitative argument before a final
conclusion can be made. In any case, this follow-up study should be done
in combination with the analysis of the large-scale structure surveys
such as 2dF and the Sloan Digital Sky Survey.

\subsubsection{Non-linear regime}

No matter what the value of $x$, there is no doubt that the non-linear
evolution of MDM must be very different from that of standard CDM. We
certainly expect MDM to eventually form mirror stars, mirror galaxies,
and so on, by analogy with ordinary matter. However, the details of this
evolution cannot be an exact parallel to that of ordinary matter. One
very important reason for this is that the mirror-helium to
mirror-hydrogen ratio from mirror primordial nucleosynthesis will be
significantly higher than its ordinary counterpart, thus affecting star
formation and evolution, which in turn will influence mirror galaxy
formation. No one has yet attempted a detailed study of this interesting
topic, apart from pointing out relatively obvious consequences such as
the faster evolution of mirror stars, and we have no progress to report
ourselves. Suffice it to say that purely observational searches for
compact mirror objects within and in the halo of our galaxy should be
pursued irrespective of the status of theoretical investigations.

The detection of early reionisation by WMAP is
potentially highly relevant for MDM. Early reionisation
implies early (ordinary) star formation, which in
turn requires sufficient power on small scales
to encourage gravitationally collapsed objects
to form. Indeed, since WMAP claims to have ruled out WDM
through this means, one might be
concerned that MDM is similarly ruled out. This is
not obviously so, however. The earliest stars must
arise from some rare large amplitude fluctuations
going non-linear. But since WDM is non-dissipative,
whereas MDM is chemically complex and dissipative,
the analogy between the two breaks down in the
non-linear regime. It is {\it a priori} possible that
the collisional damping of small scale perturbations
is compensated by the greater capacity of MDM to
clump compared to regular WDM. This is a
very interesting topic for future studies of MDM.

\section{Conclusion}
\label{sect:conclusion}

Mirror matter is a natural dark matter candidate
because it is stable. Furthermore, the mirror
matter model has aesthetic appeal through its
invariance under the full Poincar\'{e} group,
including improper Lorentz transformations.
Because the microphysics of mirror matter
is basically identical to that of ordinary matter,
the study of its cosmological implications
is well-defined, depending on a small number of
{\it a priori} free parameters
(the ratio of the relic mirror photon
temperature to that of ordinary relic photons $x$ and
the mirror baryon mass density $\Omega'_B$).

In this paper, we have looked at the linear
regime of density perturbations for mirror dark
matter in more detail than the previous major
study of Ref.\ \cite{z}. We deduced the semi-quantitative
features of the processed power
spectrum for MDM, compared it to standard CDM and HDM
and explained the origin of the differences.
A MDM power spectrum is characterised by a peak
at the scale $\lambda_{max}$ which is itself
a function of $x$ and $\Omega'_B$. The first
structures to form will have size $\lambda_{max}$.
Requiring this scale to be sub-galactic implies
that $x$ should be less than about $0.2$.
Because bottom-up structure formation is
favoured over top-down, we conclude that this is
in fact the favoured range for the temperature ratio.

It is encouraging that MDM can behave very
similarly to standard CDM for reasonable values
of $x$ and $\Omega'_B$. However, observational
differences hopefully also exist and should be used to
discriminate between the two candidates.
Future Lyman-$\alpha$ forest data probing structure at
small scales will further challenge the standard CDM
paradigm, with the prospect of uncovering a
subtle discrepancy, or of constraining models
featuring small scale damping even further.
For MDM, a constraint on small scale damping
translates into an upper bound on $x$. By using an
analogy with warm DM, we estimated an upper bound
of about $0.06$. At some
point, the need for a smaller $x$ could clash
with plausible mechanisms for baryogenesis, though
our ignorance of the correct baryogenesis mechanism
will prevent a rigorous ruling out of MDM by this means.
More optimistically, the hypothetical discovery of
small scale damping would specify the value of $x$.

The non-linear regime obviously will
differ strongly, with the generic expectation of
compact mirror objects (stars, planets, meteorites)
and structures (galaxies and so on).
The discovery of early reionisation by WMAP
potentially provides important input into the
comparison of the MDM scenario with the real universe.
Exploitation of this opportunity is not
straightforward, however, because the dissipative
and chemically complex nature of MDM acts as
a barrier to confident theoretical analysis
of the formation of the earliest stars in the MDM universe.

Finally, if mirror matter supplies all of the
non-baryonic dark matter, then existing dark matter searches for
WIMPs and axions should fail. (It is of course possible
that the universe contains both MDM and standard CDM,
even though we did not focus on that hypothesis here.)

\acknowledgments{We thank Robert Foot for comments on the manuscript.
RRV would like to thank Gary Steigman for
general discussions on early universe cosmology. He
is grateful to Ray Sawyer and the
Kavli Institute for Theoretical Physics at the University
of California at Santa Barbara for their great hospitality
during his attendance at the {\it Neutrinos: Data, Cosmos, and Planck
Scale} workshop. This work was supported in part by the
Australian Research Council and in part by the National
Science Foundation under Grant No.\ PHY99-07949.}

\end{document}